\documentclass[prb,aps,amssymb,twocolumn, superscriptaddress]{revtex4-2}
\usepackage{amsmath}
\usepackage{amsthm}
\usepackage{listings}
\usepackage{braket}
\usepackage[dvipdfmx]{graphicx}
\usepackage{multirow}
\usepackage{hyperref}
\usepackage{psfrag}
\usepackage{comment}
\usepackage{bm}
\usepackage{color}
\usepackage{lineno}

\newcommand{\beq}{\begin{equation}}
	\newcommand{\eneq}{\end{equation}}
\begin{document}
	\title{Chiral magnon in ferromagnetic chiral crystals}
	
	\author{Dapeng Yao}
	\affiliation{Department of Physics, Tokyo Institute of Technology, 2-12-1 Ookayama, Meguro-ku, Tokyo 152-8551, Japan}
		
	\author{Takehito Yokoyama}
	\affiliation{Department of Physics, Tokyo Institute of Technology, 2-12-1 Ookayama, Meguro-ku, Tokyo 152-8551, Japan}

\begin{abstract}
We theoretically propose chiral magnon in ferromagnetic chiral crystals. We show that the crystal chirality is imprinted in orbital angular momentum of magnons which exhibits the opposite signs for opposite chiralities of the crystal. We also show that a finite magnon orbital angular momentum can be induced by a temperature gradient which is a magnonic analogue of the Edelstein effect.
\end{abstract}
	
\maketitle

%{\it Introduction.}---
%\section{Introduction}
Chirality is a fundamental property of an object not identical to its mirror image by breaking of reflection or inversion symmetries. Chirality of a lattice structure, which has either a right-handed or left-handed helix, not only leads to the chirality-induced spin selectivity effect~\cite{Ray1999,Gohler2011,Naaman2012,Naaman2015} on the spin degrees of freedom, but also induces an orbital magnetization by an electric current~\cite{Yoda2015,Yoda2018}.
 Crystals with chirality exist in nature such as tellurium or selenium. Due to the low crystallographic symmetries, electricity and magnetism can be coupled in these chiral materials, leading to novel orbital effects. For instance, a current-induced orbital magnetization has been measured in nonmagnetic elemental tellurium~\cite{Furukawa2017,Furukawa2021}, and an hedgehog orbital texture can be found in $p$-type tellurium~\cite{Maruggi2023}.

Chiral phonons characterized by circular motions of ions carry phonon angular momentum, and they possess chirality due to the low-symmetry crystal structure~\cite{Zhang2014,Zhang2015,HZhu2018,Zhang2022}. Such circular rotations also emerge in a three-dimensional (3D) chiral crystal and propagate along the screw chain~\cite{Ishito2023}. 
Moreover, various unconventional physics related to phonon angular momentum have been investigated, such as couplings between chiral phonons, electrons, and magnons~\cite{Juraschek2019,Ren2021,Xiong2022,Hamada2020,Yao2024,Yao2022}, the phonon Edelstein effect~\cite{Hamada2018}, the phonon rotoelectric effect~\cite{HamadaPRB2020}, the chiral phonon diode effect~\cite{HChen2022}, and  chiral phonon-induced spin current~\cite{Kim2023,Li2024,Yao2024APL}.	

Similar to orbital effects of electrons or phonons, magnons as magnetically collective excitations show an orbital angular momentum, which results in a macroscopic orbital magnetization~\cite{Neumann2020}. In the recent works, the orbital angular momentum (OAM) of magnons has been explicitly formulated in collinear magnets~\cite{Fishman2022,Fishman_review_2022,Fishman2023}.
Different from spin magnetization as the projection of spins onto the quantization axis, OAM of magnons derives from the rotation of the perpendicular components of spinsin analogy with phonon angular momentum~\cite{Fishman2022,Fishman_review_2022} .
Recent works on magnon OAM include OAM of the twisted magnonic beams~\cite{Jia2019}, generation of magnon OAM by a skyrmion-textured domain wall in a ferromagnetic nanotube~\cite{Lee2022}, and intrinsic magnon orbital Hall effect\cite{Go}. 
 
In this Letter, considering the interplay between exchange interactions and a chiral crystal structure, we introduce a spin model with a chiral exchange interaction in a 3D chiral crystal with chirality.
Then, we theoretically propose a chiral magnon with an OAM. The crystal chirality is imprinted in OAM of chiral magnons.
In the presence of time-reversal symmetry, the total magnon OAM as a summation over the whole Brillouin zone vanishes in equilibrium because the magnon OAM of each mode is an odd function of the wavevector $\bm k$  by time-reversal symmetry~\cite{Fishman2022,Fishman_review_2022,Fishman2023}. Nevertheless, we show that a finite magnon OAM can be induced by a temperature gradient, leading to a magnon orbital Edelstein effect (MOEE). This effect is an orbital version of the magnonic analog of the Edelstein effect~\cite{BLi2020}, and analog of the Edelstein effect in electric\cite{Edelstein,Yoda2015,Yoda2018} or phononic~\cite{Hamada2018} systems.

\begin{figure}[htb]
    %\begin{figure*}[htb]
    \begin{center}
    \includegraphics[clip,width=8cm]{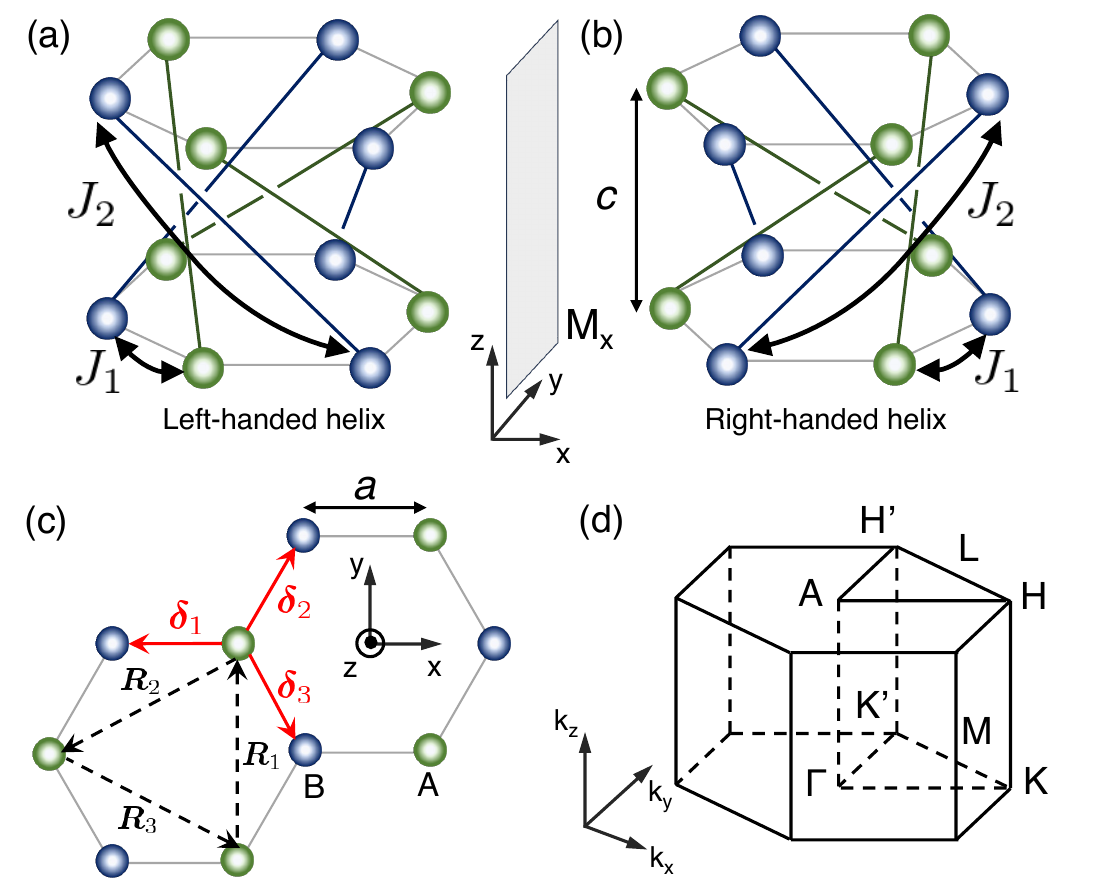}
    \end{center}
    \caption{(Color online)
    Chiral crystals composed of a stacked honeycomb lattice with (a)
    left-handed helix and (b) right-handed helix.
    They can be reflected to each other by the mirror operation $M_x$ with respect to $yz$ plane. Lattice constant along the $z$ direction is denoted by $c$.
    (c) 2D honeycomb layer. Solid arrows represent the vectors connecting the nearest-neighboring sites: $\bm \delta_1=a(-1,0)$, $\bm \delta_2=a(1/2,\sqrt{3}/2)$, and $\bm \delta_3=a(1/2,-\sqrt{3}/2)$ with $a$ being the bond length. 
    Dashed arrow denote the vectors connecting the next nearest-neighboring sites:
    $\bm R_1=a(0,\sqrt{3})$, $\bm R_2=a(-3/2,-\sqrt{3}/2)$, and $\bm R_3=a(3/2,-\sqrt{3}/2)$.
    (d) First Brillouin zone of the chiral crystal.}
     \label{struct}
     \end{figure}

%{\it Model.}---
%\section{Model} 
In order to combine the crystal chirality and ferromagnetism, we consider a 3D chiral crystal structure composed of infinitely stacked two-dimensional (2D) honeycomb lattice layers as shown in Fig.~\ref{struct}.\cite{Yoda2015,Yoda2018} The crystal structures have two distinguishable helices with the left-handed helix in Fig.~\ref{struct}(a) and the right-handed helix in Fig.~\ref{struct}(b), and they can be changed into each other by the mirror reflection $M_x$ with respect to the $yz$ plane. 
Here, we discuss the case of the right-handed helix as an example. The detailed description of these two crystal structures is included in the Supplemental Materials~\cite{SM}. For the chiral crystal with the right-handed helix, we label the nearest-neighboring vectors as $\bm \delta_i~(i=1,2,3)$ and the next nearest-neighboring vectors as $\bm R_i~(i=1,2,3)$ as shown in Fig.~\ref{struct}(c). Next, we consider the vectors connecting the atoms in the same sublattice between two layers which correpond to helical spin interactions. 
In the right-handed helix, the interlayer vectors between A sublattices are $\pm(\bm R_i+c\hat{\bm z})$, and those between B sublattices are $\pm(-\bm R_i+c\hat{\bm z})$ as shown in Fig.~\ref{struct}(b).

	% Fig.1	
     %\end{figure*}		
			
We now introduce a ferromagnetic Heisenberg spin Hamiltonian with exchange interactions for the chiral crystals, which is given by
\begin{align}\label{spinHamil0}
H=-J_1\sum_{\braket{ij}}\bm S_i\cdot\bm S_j-J_2\sum_{[ij]}\bm S_i\cdot\bm S_j,
\end{align}
where the first term denotes the nearest-neighboring exchange interaction within the honeycomb layers with the exchange interaction $J_1>0$, and the second term represents the chiral exchange interaction between the two layers with exchange parameter $J_2>0$ as shown in Figs.~\ref{struct}(a) and (b).
Then we introduce the ladder operators $S_i^{\pm}=S_i^x\pm iS_i^y$, and by means of the Holstein-Primakoff transformation~\cite{HP1940}: $S_i^z=S-a_i^{\dagger}a_i$, $S_i^{+}\approx\sqrt{2S}a_i$, and $S_i^{-}\approx\sqrt{2S}a_i^{\dagger}$ with the magnitude of spin $S$, the spin Hamiltonian in Eq.~(\ref{spinHamil0}) can be expressed in a bosonic formulation as
	\begin{align}
	H=&-J_1S\sum_{\braket{ij}}\left(a_i^{\dagger}a_j+\rm{h.c}\right)-J_2S\sum_{[ij]}\left(a_i^{\dagger}a_j+\rm{h.c}\right) \\ \nonumber
	  &+3(J_1+2J_2)S\sum_ia_i^{\dagger}a_i,   
    \end{align}
	with $a_i^{\dagger}(a_i)$ denoting the creation (annihilation) operator of magnons. Here, we show the spin Hamiltonian for the right-handed helix as an example, and that for the left-handed helix can be obtained by the mirror reflection $M_x$. The details of the derivation are given in the Supplemental Materials~\cite{SM}. For the right-handed helix, the Hamiltonian can be rewritten in terms of a quadratic Bogoliubov-de Gennes (BdG) formulation as $H=\sum_{\bm k}\bm v^{\dagger}_{\bm k}\mathcal{H}(\bm k)\bm v_{\bm k}$, where the vector operator is $\bm v_{\bm k}=(a_{\bm k,A},a_{\bm k,B},a^{\dagger}_{-\bm k,A},a^{\dagger}_{-\bm k,B})^T$, and the Bloch Hamiltonian becomes
	\begin{align}\label{spinHamil}
	\mathcal{H}(\bm k)=\frac{3}{2}(J_1+2J_2)S
	\begin{pmatrix}
	1-F_{\bm k} & -\Gamma^*_{\bm k} & 0 & 0 \\
	-\Gamma_{\bm k} & 1-G_{\bm k} & 0 & 0 \\
	0 & 0 & 1-F_{\bm k} & -\Gamma^*_{\bm k} \\
	0 & 0 & -\Gamma_{\bm k} & 1-G_{\bm k} &
 	\end{pmatrix}, 
	\end{align}
	with
	\begin{align}\label{gammak}
	\Gamma_{\bm k}=\frac{J_1}{3(J_1+2J_2)}\left\{e^{ik_xa}+2e^{-i\frac{k_xa}{2}}\cos{\left(\frac{\sqrt{3}k_ya}{2}\right)}\right\}, 
	\end{align}
	\begin{align}\label{fk} 
	F_{\bm k}=&\frac{2J_2}{3(J_1+2J_2)}\Bigg\{\cos{\left(\sqrt{3}k_ya+k_zc\right)} \\ \nonumber
	&+2\cos{\left(\frac{3k_xa}{2}\right)}\cos{\left(\frac{\sqrt{3}k_ya}{2}-k_zc\right)}\Bigg\},
	\end{align}
	\begin{align}\label{gk}
	G_{\bm k}=&\frac{2J_2}{3(J_1+2J_2)}\Bigg\{\cos{\left(\sqrt{3}k_ya-k_zc\right)} \\ \nonumber
	&+2\cos{\left(\frac{3k_xa}{2}\right)}\cos{\left(\frac{\sqrt{3}k_ya}{2}+k_zc\right)}\Bigg\}.
	\end{align}
	The corresponding magnon energies of the lower and upper modes are given by
	\begin{align}
	E_{1(2)\bm k}=3(J_1+2J_2)S\Bigg\{1-\frac{1}{2}C_{\bm k}\mp\frac{1}{2}\sqrt{B_{\bm k}^2+4|\Gamma_{\bm k}|^2}\Bigg\},
	\end{align}
	where $B_{\bm k}=G_{\bm k}-F_{\bm k}$ and $C_{\bm k}=G_{\bm k}+F_{\bm k}$.
	
	Next, we calculate the magnon OAM by following the method proposed in Refs.~\cite{Fishman2022,Fishman_review_2022,Fishman2023}. Here, we first find the inverse of the paraunitary matrix $X^{-1}(\bm k)$ determined by diagonalizing $\mathcal{H}(\bm k)\cdot N$, where $N=\sigma_z\otimes I_{2\times 2}$ is a $4\times4$ matrix with a $2\times2$ identical matrix $I_{2\times2}$ and the Pauli matrix $\sigma_z$, and introduce $\bm v_{\bm k}=X^{-1}(\bm k)\bm w_{\bm k}$ with $\bm w_{\bm k}$ denoting the vector operator in terms of the interacting Boson operators~\cite{Fishman2022,Fishman_review_2022,Fishman2023}. Under the normalization condition $X(\bm k)\cdot N\cdot X^{\dagger}(\bm k)=N$, we can obtain
	\begin{align}\label{inv_para}
	X^{-1}(\bm k)=\frac{1}{\sqrt{2}\Gamma^*_{\bm k}}
	\begin{pmatrix}
	\Gamma^*_{\bm k}K_{\bm k}^- & -\Gamma^*_{\bm k}K_{\bm k}^+ & 0 & 0 \\
	|\Gamma_{\bm k}|K_{\bm k}^+ & |\Gamma_{\bm k}|K_{\bm k}^- & 0 & 0 \\
	0 & 0 & \Gamma^*_{\bm k}K_{\bm k}^- & -\Gamma^*_{\bm k}K_{\bm k}^+ \\
	0 & 0 & |\Gamma_{\bm k}|K_{\bm k}^+ & |\Gamma_{\bm k}|K_{\bm k}^-
	\end{pmatrix},
	\end{align}
	where
	\begin{align}
	K_{\bm k}^{\pm}=\sqrt{1\pm\frac{B_{\bm k}}{\sqrt{B_{\bm k}^2+4|\Gamma_{\bm k}|^2}}}.
	\end{align}
Details of this derivation are given in Supplemental Material~\cite{SM}.

\begin{figure}[htb]
    %\begin{figure*}[htb]
    \begin{center}
    \includegraphics[clip,width=8cm]{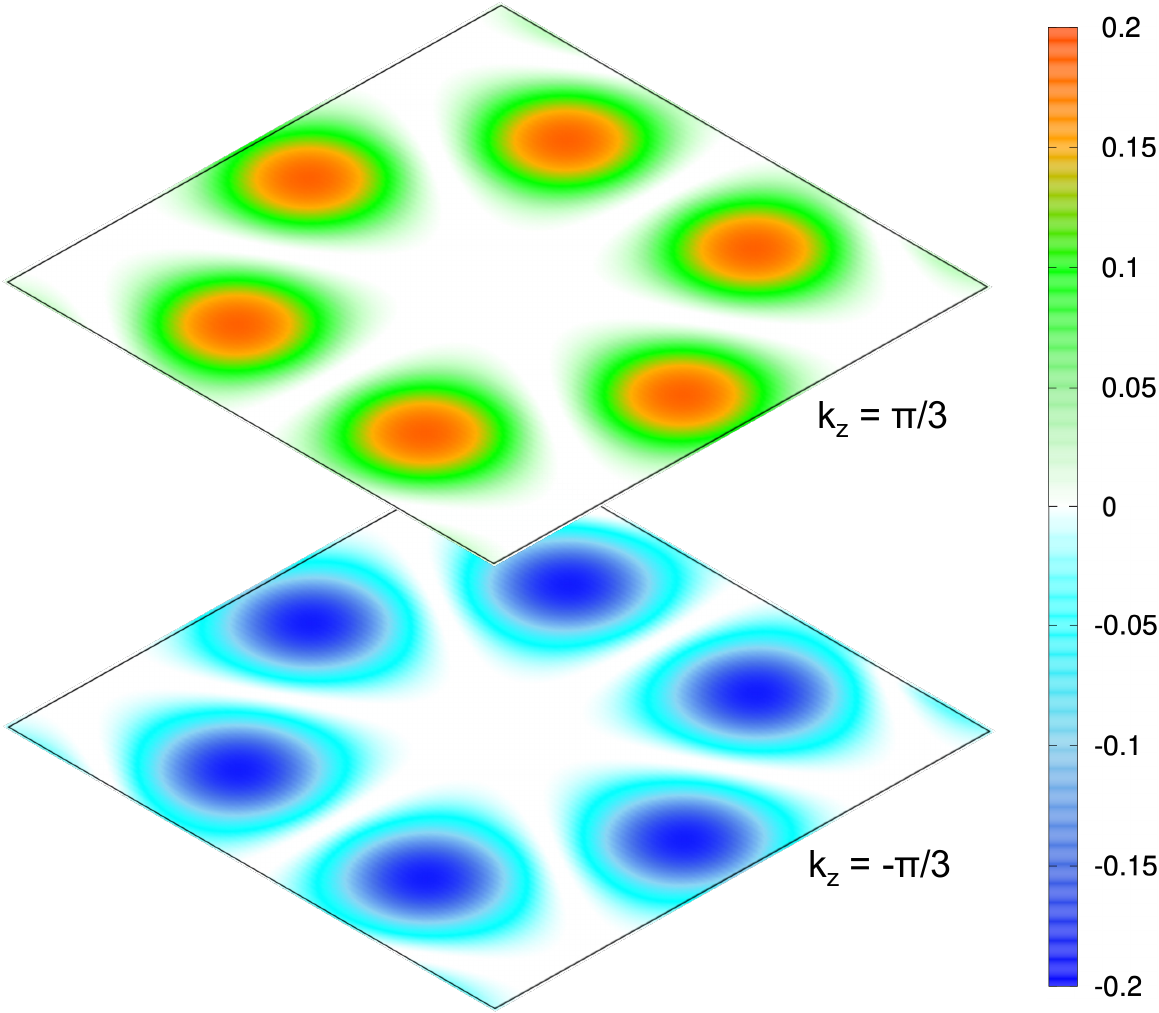}
    \end{center}
    \caption{(Color online)
    Distribution of $\Delta\mathcal{L}_{z2}(\bm k)$ in Eq.~(\ref{DLz}) in the $k_x$-$k_y$ plane. Color represents the magnon OAM in the unit of $\hbar$. The parameters are set to be $J_2=0.5J_1$.}
     \label{dist_Lz}
     \end{figure}

	 %{\it Chiral magnon with OAM.}---
	 %\section{Chiral magnon with OAM}
	 Here, the magnon OAM can be calculated by using the inverse of the paraunitary matrix $X^{-1}(\bm k)$ derived from the BdG Hamiltonian in Eq.~(\ref{spinHamil})~\cite{Fishman2022,Fishman_review_2022,Fishman2023}. The expectation value of the OAM as a function of wavevector $\bm k$  in the $n$-th eigenmode for the chiral crystal with the right-handed helix is given by
	\begin{align}\label{gen_Lzn}
	\mathcal{L}_{zn}(\bm k)=\frac{\hbar}{2}\sum_{r=1}^2\Bigg\{&X^{-1}(\bm k)_{rn}\hat{l}_{z\bm k}X^{-1}(\bm k)_{rn}^* \\ \nonumber
	&-X^{-1}(\bm k)_{r+2,n}\hat{l}_{z\bm k}X^{-1}(\bm k)_{r+2,n}^*\Bigg\},
	\end{align}
	where $\hat{l}_{z\bm k}=-i\left(\bar{k}_x\partial_{k_y}-\bar{k}_y\partial_{k_x}\right)$ represents the operator of magnon OAM with the periodic functions $\bar{k}_xa=\sin{(3k_xa/2)}\cos{(\sqrt{3}k_ya/2)}$ and $\sqrt{3}\bar{k}_ya=\sin{(\sqrt{3}k_ya/2)}\cos{(3k_xa/2)}+\sin{(\sqrt{3}k_ya)}$ within the 2D honeycomb layers to guarantee the periodicity of magnon OAM on a discrete lattice~\cite{Fishman2022,Fishman_review_2022}.
	Then, the magnon OAM in Eq.~(\ref{gen_Lzn}) can be further expressed as
	\begin{align}\label{Lz_3D}
    \mathcal{L}_{z1(2)}(\bm k)=\frac{\hbar}{4}\left(1\pm\frac{B_{\bm k}}{\sqrt{B_{\bm k}^2+4|\Gamma_{\bm k}|^2}}\right)\frac{\Gamma_{\bm k}}{|\Gamma_{\bm k}|}\hat{l}_{z\bm k}\frac{\Gamma^*_{\bm k}}{|\Gamma_{\bm k}|},
	\end{align}
    for the lower and upper bands by means of Eq.~(\ref{inv_para}). Here, we notice that the first term 
    \begin{align}\label{L0}
    \mathcal{L}_0(\bm k)=\frac{\hbar}{4}\frac{\Gamma_{\bm k}}{|\Gamma_{\bm k}|}\hat{l}_{z\bm k}\frac{\Gamma^*_{\bm k}}{|\Gamma_{\bm k}|}
    \end{align}
    denotes the contribution from the 2D honeycomb layers, which is a function of wavenumbers $k_x$ and $k_y$, and has been thoroughly investigated by the previous studies~\cite{Fishman2022,Fishman_review_2022,Fishman2023}. On the other hand, the second term 
    \begin{align}\label{DLz}
	\Delta\mathcal{L}_{z1(2)}(\bm k)=\pm\frac{\hbar}{4}\frac{B_{\bm k}}{\sqrt{B_{\bm k}^2+4|\Gamma_{\bm k}|^2}}\frac{\Gamma_{\bm k}}{|\Gamma_{\bm k}|}\hat{l}_{z\bm k}\frac{\Gamma^*_{\bm k}}{|\Gamma_{\bm k}|},
	\end{align}
	which comes from the chiral exchange interaction is a function of wavenumbers $k_x$, $k_y$ and $k_z$, and plays an essential role in chiral magnons. Figure~\ref{dist_Lz} show the distribution of $\Delta\mathcal{L}_{z2}(\bm k)$ in the $k_x$-$k_y$ momentum space for $k_z=\pm\pi/3$.    
We see that $\Delta\mathcal{L}_{z2}(\bm k)$ is six fold symmetric around the $z$-axis and odd in $k_z$.

    % Fig.2	
     %\end{figure*}
    
\begin{figure}[htb]
    %\begin{figure*}[htb]
    \begin{center}
    \includegraphics[clip,width=8cm]{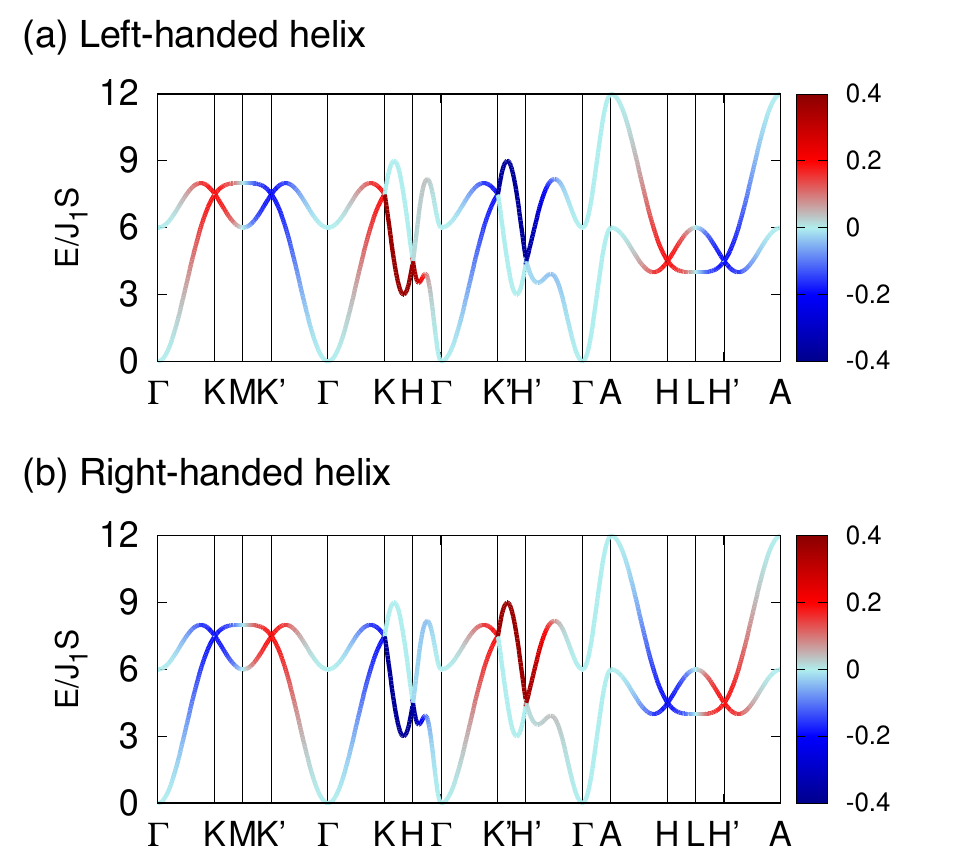}
    \end{center}
    \caption{(Color online)
    Magnon spectra for the (a) left-handed helix and (b) right-handed helix with the color
    representing the magnon OAM $\mathcal{L}_{z1(2)}$ in Eq.~(\ref{Lz_3D}) in the unit of $\hbar$. The parameters are set to be $J_2=0.5J_1$.  
     }
     \label{band_OAM}
     \end{figure}

     The numerical results of the $\bm k$-dependent magnon OAM distributions $\mathcal{L}_{zn}(\bm k)$ in Eq.~(\ref{Lz_3D}) of each magnon band along the high-symmetry points for the left-handed and right-handed helices are shown in Fig.~\ref{band_OAM}(a) and Fig.~\ref{band_OAM}(b), respectively. 
     One can clearly see that the magnon OAM has the opposite signs between the chiral crystals with the opposite helices. Here, for a given wavevector $\bm k$, the opposite signs of the magnon OAMs mean that the rotations of the spins around the $z$ axis are in the opposite directions for different helices. Therefore, crystal chirality is imprinted in magnon OAM in chiral crystals. In addition to magnon OAM, spin angular momentum plays an important role in ferromagnets. However, here the spin angular momentum has no difference between the two chiral structures because they can be reflected to each other via the mirror operation $M_x$:
    $U\mathcal{H}^R(k_x,k_y,k_z)U^{-1}=\mathcal{H}^L(-k_x,k_y,k_z)$,
    where $\mathcal{H}^R(\bm k)$ and $\mathcal{H}^L(\bm k)$ are the Bloch spin Hamiltonians for the right-handed and left-handed helices, respectively, where the unitary matrix is given by $U=I_{2\times2}\otimes\sigma_x$ in our model. 
    Thus, the expectation values of the spin along the quantization axis satisfy $\braket{S^L_z}=\braket{S^R_z}$.

  In equilibrium, the total magnon OAM for the $z$ component per volume $V$ is given by
	\begin{align}\label{Li_orb}
	L_z^{\rm{orb}}=\frac{1}{V}\sum_{\bm k,n=1,2}\mathcal{L}_{zn}(\bm k)\left\{2f_0(E_{n\bm k})+1\right\},
	\end{align}
	where $f_0(E_{n\bm k})=1/\left(e^{E_{n\bm k}/k_BT}-1\right)$ is the Bose-Einstein distribution function~\cite{Fishman2023}. In this case, the total magnon OAM vanishes because time-reversal symmetry requires $X^{-1}(-\bm k)=X^{-1}(\bm k)^*$, resulting in $\mathcal{L}_{zn}(\bm k)=-\mathcal{L}_{zn}(-\bm k)$~\cite{Fishman2022,Fishman_review_2022,Fishman2023}. Nevertheless, a finite magnon OAM can be generated by a temperature gradient because the magnon distribution can become out of equilibrium.

    % Fig.3	
     %\end{figure*}
	
	%{\it Magnon orbital Edelstein effect.}---
	%\section{Magnon orbital Edelstein effect}	
	In analogy with the Edelstein effect in phononic systems~\cite{Hamada2018,HChen2022}, here we theoretically show that a chiral magnon carrying an OAM can be driven by a temperature gradient in chiral crystals. This phenomenon can be considered as MOEE.
     In general, a non-equilibrium magnon distribution can be obtained when a temperature gradient $\partial T/\partial x_i$ is applied. By using the Boltzmann  equation with the relaxation time approximation, the distribution function of magnons in nonequilibrium is given by
	\begin{align}
	f_{n\bm k}=f_0(E_{n\bm k})-\tau v_{n\bm k,i}\frac{\partial f_0}{\partial T}\frac{\partial T}{\partial x_i},
	\end{align}
	where $\tau$ represents the magnon relaxation time and $v_{n\bm k,i}=\partial E_{n\bm k}/\hbar\partial k_i$ is the group velocity of the $n$-th mode of magnons.
	Accordingly, the magnon OAM per unit volume generated by the temperature gradient becomes
	\begin{align}\label{outeqL_orb}
	L_i^{\rm{orb}}=-\frac{2\tau}{\hbar V}\sum_{n\bm k}\mathcal{L}_{in}(\bm k)\frac{\partial E_{n\bm k}}{\partial k_j}\frac{\partial f_0(E_{n\bm k})}{\partial T}\frac{\partial T}{\partial x_j}\equiv\alpha_{ij}\frac{\partial T}{\partial x_j},
	\end{align}
	where $\alpha_{ij}$ denotes the response tensor, which can be determined by the point-group symmetry. In our model, for the chiral crystals with the $D_6$ point group, the response tensor is generally represented as\cite{Hamada2018}
	\begin{align}\label{symmetry}
	\alpha_{ij}=
	\begin{pmatrix}
	\alpha_{xx} & 0 & 0 \\
	0 & \alpha_{xx} & 0 \\
	0 & 0 & \alpha_{zz}
	\end{pmatrix}.
	\end{align} 		
	% Fig.4	
	\begin{figure}[htb]
    %\begin{figure*}[htb]
    \begin{center}
    \includegraphics[clip,width=9cm]{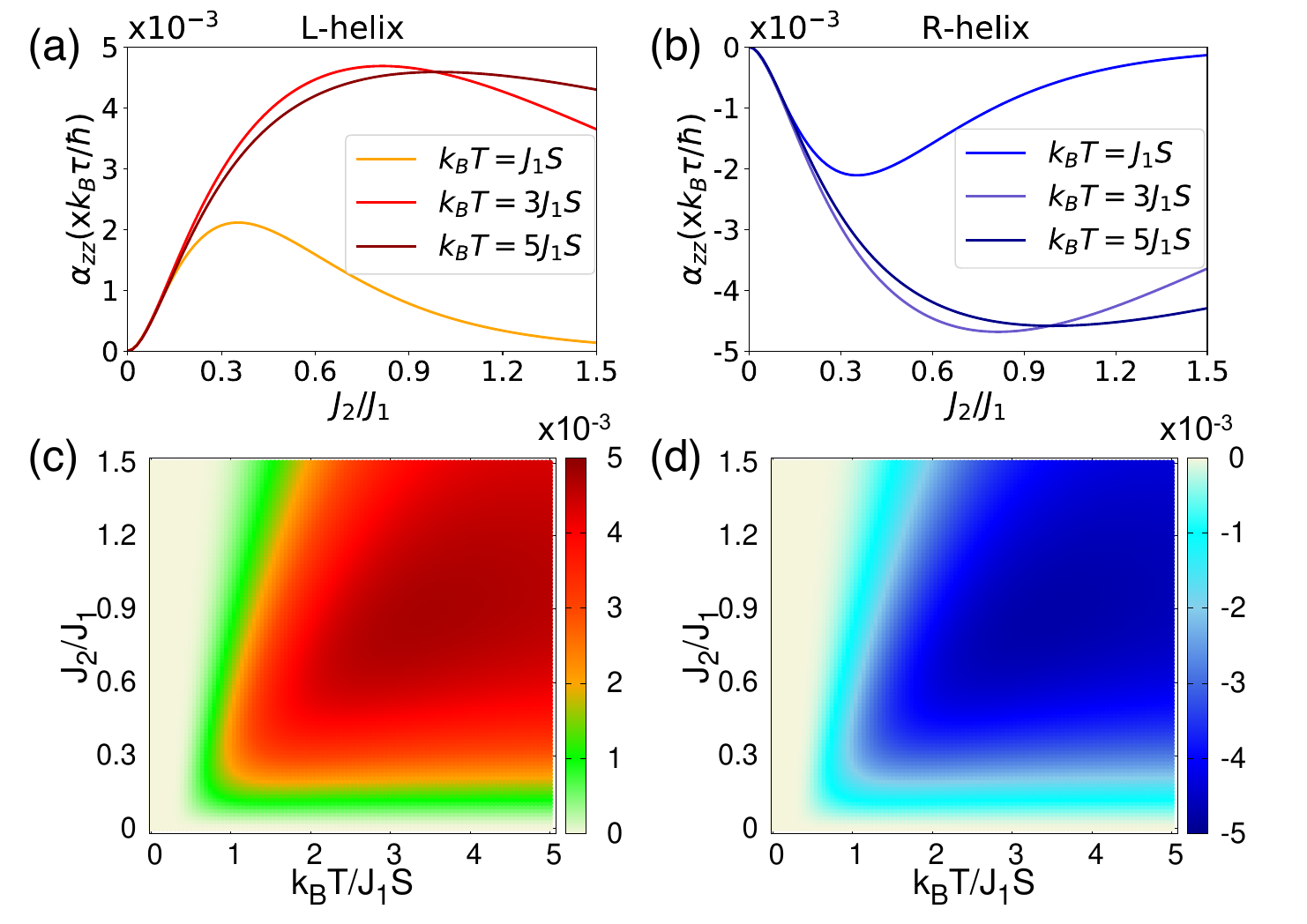}
    \end{center}
    \caption{(Color online)
    Calculated MOEE coefficient $\alpha_{zz}$ given in Eq.~(\ref{al_zz}). (a) and (b) show $\alpha_{zz}$ versus chiral exchange interaction $J_2$ with different temperatures for the left-handed and right-handed helices, respectively. (c) and (d) show $\alpha_{zz}$ as a function of $k_BT/J_1S$ and $J_2/J_1$ for the left-handed and right-handed helices, respectively.
     }
     \label{alpha}
     \end{figure}
     %\end{figure*}
Therefore, in the case of the chiral crystals shown in Figs.~\ref{struct}(a) and (b), when the temperature gradient is applied along the $z$ direction, we can obtain  a finite  $z$ component of magnon OAM: $L_z^{\rm{orb}}=\alpha_{zz}\partial_zT$, with the temperature gradient $\partial_zT$ and the response coeffcient $\alpha_{zz}$ of the MOEE.
	Here the $\bm k$-dependent magnon OAM satisfies $\mathcal{L}_{zn}(k_x,k_y,k_z)=\mathcal{L}_0(k_x,k_y)+\Delta\mathcal{L}_{zn}(k_x,k_y,k_z)$ as given in Eqs.~(\ref{Lz_3D}), (\ref{L0}), and (\ref{DLz}). We notice that the summation in Eq.~(\ref{outeqL_orb}) for the first term $\mathcal{L}_0(k_x,k_y)$ becomes zero due to the time-reversal symmetry.  $\mathcal{L}_0(k_x,k_y)$ is independent of $k_z$, and corresponds to the spin waves only propagating within the 2D honeycomb layer for a given wavevector $\bm k$. 
	As a result, only $\Delta\mathcal{L}_{zn}(k_x,k_y,k_z)$ contributes to the MOEE coefficient $\alpha_{zz}$, and $\alpha_{zz}$ eventually becomes
	\begin{align}\label{al_zz}
	\alpha_{zz}=-\frac{2\tau}{\hbar V}\sum_{\bm k,n=1,2}\Delta\mathcal{L}_{zn}(\bm k)\frac{\partial E_{n\bm k}}{\partial k_z}\frac{\partial f_0(E_{n\bm k})}{\partial T},
	\end{align}
	for the chiral crystals. Figure~\ref{alpha} shows the numerical results of the MOEE coefficient $\alpha_{zz}$ based on Eq.~(\ref{al_zz}). We see that the MOEE coefficient $\alpha_{zz}$ is not monotonic in $T$ and $J_2$. The sign of $\alpha_{zz}$ becomes opposite between the crystal structures with the opposite helices because $\Delta\mathcal{L}_{zn}(k_x,k_y,k_z)$ changes the sign when the helix is changed into its mirror image as shown in Fig.~\ref{band_OAM}. Here we notice that the chiral exchange interaction in Eq.~(\ref{spinHamil0}) contributes to the chirality of magnon OAM.
	
Let us estimate the size of the magnon OAM generated by a temperature gradient. Here, we set the parameters $J_2=0.5J_1$ with $J_1=1$meV, $k_BT=2.5J_1S$, and a lattice constance $a=10^{-9}$m. The MOEE coefficient is estimated as $\alpha_{zz}\sim 10^{26}\hbar\times[\tau/(1$s$)]$m$^{-2}$K$^{-1}$. We use the relaxation time of magnon $\tau=10^{-10}$s and a temperature gradient $\partial_zT=10$K/mm~\cite{BLi2020}. Then, we can estimate the magnon OAM generated along the $z$ direction with the size of $10^{14}\hbar/$cm$^{3}$, which is of the same order as the magnon Edelstein effect due to the spin angular momentum~\cite{BLi2020} and experimentally observable.
Here, we present an experimental proposal for the magnon OAM detection. The magnon OAM can have a topological nature~\cite{Fishman2022,Fishman_review_2022,Fishman2023}, i.e., $\mathcal{L}_{zn}(\bm k)$ can be rewritten by the cross product of the momentum $\bm k$ and the Berry connection $\bm{\mathcal{A}}_n(\bm k)$ as
\begin{align}
\mathcal{L}_{zn}(\bm k)=-\frac{\hbar}{2}\Big\{\bm k\times\bm{\mathcal{A}}_n(\bm k)\Big\}\cdot\hat{\bm z},
\end{align}
with
\begin{align}
\bm{\mathcal{A}}_{n}(\bm k)\equiv i\sum_{r=1}^{M}\Bigg[&X^{-1}_{rn}(\bm k)^*\frac{\partial X^{-1}_{rn}(\bm k)}{\partial\bm k} \nonumber\\
&-X^{-1}_{r+M,n}(\bm k)^*\frac{\partial X^{-1}_{r+M,n}(\bm k)}{\partial\bm k}\Bigg].
\end{align}
Because the Berry connection shows sharp peaks near the valley points $\bm K$ and $\bm K'$ and plummets suddenly  away from these points~\cite{SM}, here only the contributions from the valley points to the Berry connection are considered. Then, the MOEE coefficient can be approximately rewritten by
\begin{align}
\alpha_{zz}\approx~2\tau\left[\bm K\times\Delta\bm P_{\text{mag}}\right]_z,
\end{align}
where
\begin{align}
\Delta\bm P_{\text{mag}}=\sum_n\int\frac{d\bm k}{(2\pi)^3}\Delta\bm{\mathcal{A}}_{n}(\bm k)\frac{\partial E_{n\bm k}}{\partial k_z}\frac{\partial f_0(E_{n\bm k})}{\partial T},
\end{align}
can be considered as the magnon ``polarization" given in terms of the Berry connection $\Delta\bm{\mathcal{A}}_{n}(\bm k)$ corresponding to $\Delta\mathcal{L}_{zn}(\bm k)$ in Eq.~(\ref{DLz}).
Thus, under the temperature gradient along the screw axis, the generated magnon OAM accompanies the magnon current perpendicular to the screw axis because the change of magnon ``polarization" $\Delta\bm P_{\text{mag}}$ accompanies  the change of magnon position, in analogy to electric polarization~\cite{King-Smith1993}.
Since magnon (spin) current can be measured via the inverse spin Hall effect~\cite{Uchida2010}, the MOEE can be measured through inverse spin Hall effect.
We provide more details about experimental proposal in the Supplemental Material~\cite{SM}.

To summarize, we have theoretically proposed chiral magnon with OAM, which exhibits a chirality in a ferromagnetic chiral crystal due to the lack of inversion and mirror symmetries. A finite magnon orbital angular momentum can be generated by a temperature gradient, leading to MOEE. 
Therefore, chiral crystals provide a platform for detecting the magnon OAM which can be measured by inverse spin Hall effect.
We propose a candidate material Cu$_2$OSeO$_3$, which is a magnetic insulator with a chiral structure~\cite{Meunier1976,Seki2012}. Because temperature gradient leads to a redistribution of electrons in metals, this insulating material is suitable for future measurement.

D.Y. was supported by JSPS KAKENHI Grants No.~JP23KJ0926. T.Y. was supported by JSPS KAKENHI Grant No.~JP30578216.

%\bibliography{main.bib}

\end{document}

\title{Supplemental Material for ``Chiral magnon in ferromagnetic chiral crystals"}

\author{Dapeng Yao}
\affiliation{Department of Physics, Tokyo Institute of Technology, 2-12-1 Ookayama, Meguro-ku, Tokyo 152-8551, Japan}

\author{Takehito Yokoyama}
\affiliation{Department of Physics, Tokyo Institute of Technology, 2-12-1 Ookayama, Meguro-ku, Tokyo 152-8551, Japan}

\maketitle

%\appendix
%\maketitle

%\author{Name}
%\affiliation{affiliation}
\renewcommand{\theequation}{S\arabic{equation}}
\setcounter{equation}{0}
\renewcommand{\tablename}{Table S}
\setcounter{table}{0}
\renewcommand{\thefigure}{S\arabic{figure}}
\setcounter{figure}{0}

\renewcommand{\thesection}{Supplemental Material~\Roman{section}}
\setcounter{section}{0}

\onecolumngrid

%\tableofcontents

%\clearpage

\section{Derivation of the Bloch spin Hamiltonian}

\begin{figure}[htb]
    %\begin{figure*}[htb]
    \begin{center}
    \includegraphics[clip,width=8cm]{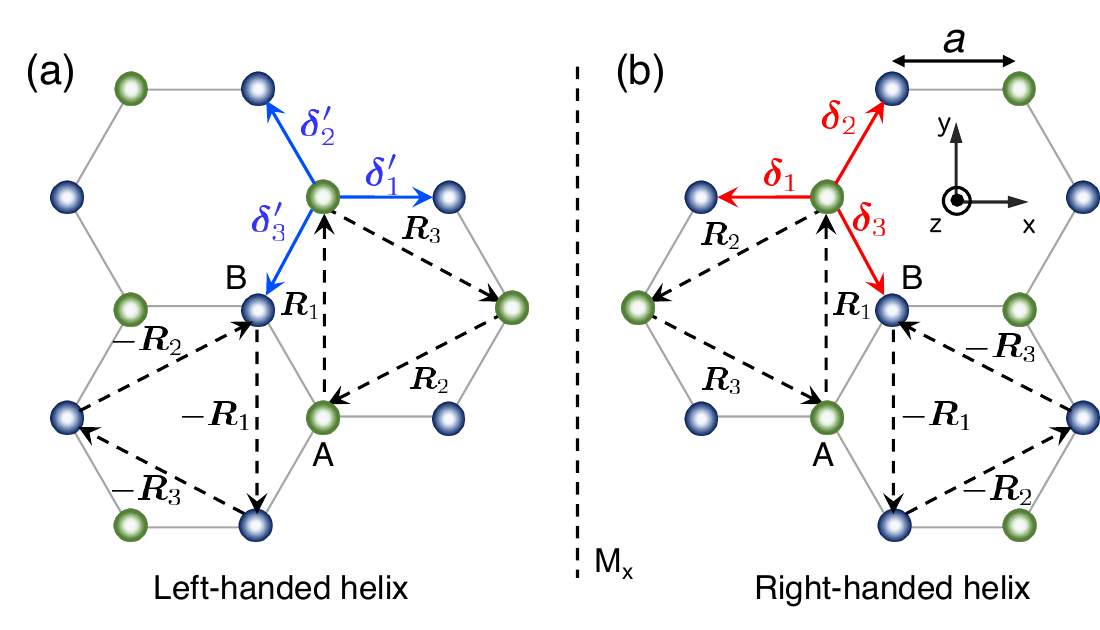}
    \end{center}
    \caption{Schematics of two chiral structures with (a) the left-handed helix and (b) the right-handed helix. The vectors connecting the nearest-neighbor sites in (a) the left-handed helix are $\bm \delta_1'=a(1,0)$, $\bm \delta_2'=a(-1/2,\sqrt{3}/2)$, and $\bm \delta_3'=a(-1/2,-\sqrt{3}/2)$, and those in (b) the right-handed helix are $\bm \delta_1=a(-1,0)$, $\bm \delta_2=a(1/2,\sqrt{3}/2)$, and $\bm \delta_3=a(1/2,-\sqrt{3}/2)$.
    }
     \label{sm_struct}
     \end{figure}

Here we show the derivation of the spin Hamiltonian Eq.~(2) in the main text for the chiral structures with the right-handed and left-handed helices. In fact, the choice of the unit cell is not unique, and here in order to discuss the results clearly, the unit cells in the crystals with the opposite chirality are set to be mirror image of each others as shown in Fig.~\ref{sm_struct}. Here, $\bm \delta_i'~(i=1,2,3)$ represents the vector connecting the nearest-neighbor sites in Fig.~\ref{sm_struct}(a) for the left-handed helix, and $\bm \delta_i~(i=1,2,3)$ represents that in Fig.~\ref{sm_struct}(b) for the right-handed helix.
The chiral structures shown in Figs.~1(a)(b) in the main text consist of honeycomb-lattice layers, stacked along the $z$ axis. Vectors for the chiral exchange interactions between two atoms are  combinations of the vectors connecting the next-nearest-neighbor sites $\bm R_i~(i=1,2,3)$ and the translation of $c$ along the $z$ axis. The interlayer vectors between A sublattices are $\pm\left(\bm R_i+c\hat{z}\right)$, and those between B sublattices are $\pm\left(-\bm R_i+c\hat{z}\right)$ in the right-handed coordinate system.

% Fig.S1	
     %\end{figure*}	

Then, the first term of the spin Hamiltonian in Eq.~(2) for the structure with the left-handed helix can be written as
\begin{align}
H_1^{\rm{L}}=-J_1S\sum_{\bm r}\sum_{i=1}^3 a_{B,\bm r+\bm \delta_i'}^{\dagger}a_{A,\bm r}+\text{h.c.}
=\sum_{\bm k}
\begin{pmatrix}
a^{\dagger}_{A,\bm k} & a^{\dagger}_{B,\bm k}
\end{pmatrix}
\begin{pmatrix}
0 & -\gamma_{\bm k} \\
-\gamma^*_{\bm k} & 0
\end{pmatrix}
\begin{pmatrix}
a_{A,\bm k} \\
a_{B,\bm k}
\end{pmatrix},
\end{align}
and that for the structure with the right-handed helix can be written as
\begin{align}
H_1^{\rm{R}}=-J_1S\sum_{\bm r}\sum_{i=1}^3 a_{B,\bm r+\bm \delta_i}^{\dagger}a_{A,\bm r}+\text{h.c.} 
=\sum_{\bm k}
\begin{pmatrix}
a^{\dagger}_{A,\bm k} & a^{\dagger}_{B,\bm k}
\end{pmatrix}
\begin{pmatrix}
0 & -\gamma^*_{\bm k} \\
-\gamma_{\bm k} & 0
\end{pmatrix}
\begin{pmatrix}
a_{A,\bm k} \\
a_{B,\bm k}
\end{pmatrix},
\end{align}
where
\begin{align}
\gamma_{\bm k}=J_1S\sum_{i=1}^3e^{i\bm k\cdot\bm \delta'_i}=J_1S\left(\sum_{i=1}^3e^{i\bm k\cdot\bm \delta_i}\right)^* 
=J_1S\left\{e^{ik_xa}+e^{-i(k_xa+\sqrt{3}k_y)/2}+e^{-i(k_xa-\sqrt{3}k_y)/2}\right\},
\end{align}
with $\bm \delta_1=-\bm \delta'_1$, $\bm \delta_2=-\bm \delta'_3$, and $\bm \delta_3=-\bm \delta'_2$. Next, the second term of the spin Hamiltonian in Eq.~(2) for the left-handed helix and that for right-handed helix are the same if we take the vectors in Fig.~\ref{sm_struct}, which is given by
\begin{align}
H_2=&-J_2S\sum_{\bm r}\sum_{i=1}^3a^{\dagger}_{A,\bm r+\bm R_i+c\hat{z}}a_{A,\bm r}+a^{\dagger}_{A,\bm r-\bm R_i-c\hat{z}}a_{A,\bm r} \nonumber 
-J_2S\sum_{\bm r}\sum_{i=1}^3a^{\dagger}_{B,\bm r-\bm R_i+c\hat{z}}a_{B,\bm r}+a^{\dagger}_{B,\bm r+\bm R_i-c\hat{z}}a_{B,\bm r} \nonumber \\
=&-J_2S\sum_{\bm k}
\begin{pmatrix}
a^{\dagger}_{A,\bm k} & a^{\dagger}_{B,\bm k}
\end{pmatrix}
\begin{pmatrix}
h^A_{\bm k} & 0 \\
0 & h^B_{\bm k}
\end{pmatrix}
\begin{pmatrix}
a_{A,\bm k} \\
a_{B,\bm k}
\end{pmatrix},
\end{align}
with
\begin{align}
h^A_{\bm k}=&\cos\left(\sqrt{3}k_ya+k_zc\right)+\cos\left(-\frac{3k_xa}{2}-\frac{\sqrt{3}k_ya}{2}+k_zc\right)+\cos\left(\frac{3k_xa}{2}-\frac{\sqrt{3}k_ya}{2}+k_zc\right) \\
h^B_{\bm k}=&\cos\left(\sqrt{3}k_ya-k_zc\right)+\cos\left(-\frac{3k_xa}{2}-\frac{\sqrt{3}k_ya}{2}-k_zc\right)+\cos\left(\frac{3k_xa}{2}-\frac{\sqrt{3}k_ya}{2}-k_zc\right)
\end{align}
The last term of the spin Hamiltonian in Eq.~(2) can be written as
\begin{align}
H_3=3(J_1+2J_2)S\sum_{\bm k}
\begin{pmatrix}
a^{\dagger}_{A,\bm k} & a^{\dagger}_{B,\bm k}
\end{pmatrix}
\begin{pmatrix}
1 & 0 \\
0 & 1
\end{pmatrix}
\begin{pmatrix}
a_{A,\bm k} \\
a_{B,\bm k}
\end{pmatrix}.
\end{align}
Therefore, we can write down the total spin Hamiltonian in terms of a quadratic Bogoliubov de Gennes form as $H^{\alpha}=\sum_{\bm k}\bm v^{\dagger}_{\bm k}\mathcal{H}^{\alpha}(\bm k)\bm v_{\bm k}$ with $\alpha=\rm{R/L}$, the vector operator $\bm v_{\bm k}=\left(a^{\dagger}_{A,\bm k}~a^{\dagger}_{B,\bm k}~a_{A,-\bm k}~a_{B,-\bm k}\right)^{T}$, the Bloch Hamiltonian
\begin{align}\label{SM_left}
\mathcal{H}^{\rm L}(\bm k)=\frac{3}{2}(J_1+2J_2)S
\begin{pmatrix}
1-F_{\bm k} & -\Gamma_{\bm k} & 0 & 0 \\
-\Gamma^*_{\bm k} & 1-G_{\bm k} & 0 & 0 \\
0 & 0 & 1-F_{\bm k} & -\Gamma_{\bm k} \\
0 & 0 & -\Gamma^*_{\bm k} & 1-G_{\bm k}
\end{pmatrix}
\end{align}
for the left-handed helix and that
\begin{align}\label{SM_right}
\mathcal{H}^{\rm R}(\bm k)=\frac{3}{2}(J_1+2J_2)S
\begin{pmatrix}
1-F_{\bm k} & -\Gamma^*_{\bm k} & 0 & 0 \\
-\Gamma_{\bm k} & 1-G_{\bm k} & 0 & 0 \\
0 & 0 & 1-F_{\bm k} & -\Gamma^*_{\bm k} \\
0 & 0 & -\Gamma_{\bm k} & 1-G_{\bm k}
\end{pmatrix},
\end{align}
for the right-handed helix, where 
\begin{align}\label{gammak}
	\Gamma_{\bm k}=\frac{J_1}{3(J_1+2J_2)}\left\{e^{ik_xa}+2e^{-i\frac{k_xa}{2}}\cos{\left(\frac{\sqrt{3}k_ya}{2}\right)}\right\}, 
	\end{align}
	\begin{align}\label{fk} 
	F_{\bm k}=\frac{2J_2}{3(J_1+2J_2)}\Bigg\{\cos{\left(\sqrt{3}k_ya+k_zc\right)} 
	+2\cos{\left(\frac{3k_xa}{2}\right)}\cos{\left(\frac{\sqrt{3}k_ya}{2}-k_zc\right)}\Bigg\},
	\end{align}
	\begin{align}\label{gk}
	G_{\bm k}=&\frac{2J_2}{3(J_1+2J_2)}\Bigg\{\cos{\left(\sqrt{3}k_ya-k_zc\right)} 
	+2\cos{\left(\frac{3k_xa}{2}\right)}\cos{\left(\frac{\sqrt{3}k_ya}{2}+k_zc\right)}\Bigg\}.
\end{align}

\section{Magnon orbital angular momentum}
In this section, we follow the method proposed by Fishman $et.~al.$~\cite{SM_Fishman2022} to calculate the magnon orbital angular momentum (OAM) for the ferromagnetic chiral crystal. Here we show the calculation for the right-handed helix from the Hamiltonian in Eq.~(\ref{SM_right}) as an example, and that for the left-handed helix from the Hamiltonian in Eq.~(\ref{SM_left}) can be similarly performed.
Based on the Hamiltonian in Eq.~(\ref{SM_right}), we first find that the eigenvectors of $\mathcal{H}^{\rm R}(\bm k)\cdot N$ are
\begin{align}
\bm u_1=X(\bm k)^*_{1j}=c^*_1
\begin{pmatrix}
G_{\bm k}-F_{\bm k}-\sqrt{4|\Gamma_{\bm k}|^2+(G_{\bm k}-F_{\bm k})^2} & -2\Gamma_{\bm k} & 0 & 0
\end{pmatrix}, \\
\bm u_2=X(\bm k)^*_{2j}=c^*_2
\begin{pmatrix}
G_{\bm k}-F_{\bm k}+\sqrt{4|\Gamma_{\bm k}|^2+(G_{\bm k}-F_{\bm k})^2} & -2\Gamma_{\bm k} & 0 & 0
\end{pmatrix}, \\
\bm u_3=X(\bm k)^*_{3j}=c_3
\begin{pmatrix}
0 & 0 & G_{\bm k}-F_{\bm k}-\sqrt{4|\Gamma_{\bm k}|^2+(G_{\bm k}-F_{\bm k})^2} & -2\Gamma_{\bm k}
\end{pmatrix}, \\
\bm u_4=X(\bm k)^*_{4j}=c_4
\begin{pmatrix}
0 & 0 & G_{\bm k}-F_{\bm k}+\sqrt{4|\Gamma_{\bm k}|^2+(G_{\bm k}-F_{\bm k})^2} & -2\Gamma_{\bm k}
\end{pmatrix},
\end{align}
where $N=\sigma_z\otimes I_{2\times2}$. Hence, the paraunitary matrix of the Hamiltonian in Eq.~(\ref{SM_right}) can be expressed as
\begin{align}
X(\bm k)=
\begin{pmatrix}
c_1\left(B_{\bm k}-\sqrt{B_{\bm k}^2+|A_{\bm k}|^2}\right) & -c_1A^*_{\bm k} & 0 & 0 \\
c_2\left(B_{\bm k}+\sqrt{B_{\bm k}^2+|A_{\bm k}|^2}\right) & -c_2A^*_{\bm k} & 0 & 0 \\
0 & 0 & c^*_3\left(B_{\bm k}-\sqrt{B_{\bm k}^2+|A_{\bm k}|^2}\right) & -c^*_3A^*_{\bm k}  \\
0 & 0 & c^*_4\left(B_{\bm k}+\sqrt{B_{\bm k}^2+|A_{\bm k}|^2}\right) & -c^*_4A^*_{\bm k}  \\
\end{pmatrix},
\end{align}
where $A_{\bm k}=2\Gamma_{\bm k}$ and $B_{\bm k}=G_{\bm k}-F_{\bm k}$, and its inverse is
\begin{align}\label{SM_invpara}
X^{-1}(\bm k)=\frac{1}{2A_{\bm k}^*}
\begin{pmatrix}
-\frac{A_{\bm k}^*}{c_1\sqrt{B_{\bm k}^2+|A_{\bm k}|^2}} & \frac{A_{\bm k}^*}{c_2\sqrt{B_{\bm k}^2+|A_{\bm k}|^2}} & 0 & 0 \\
-\frac{B_{\bm k}+\sqrt{B_{\bm k}^2+|A_{\bm k}|^2}}{c_1\sqrt{B_{\bm k}^2+|A_{\bm k}|^2}} & \frac{B_{\bm k}-\sqrt{B_{\bm k}^2+|A_{\bm k}|^2}}{c_2\sqrt{B_{\bm k}^2+|A_{\bm k}|^2}} & 0 & 0 \\
0 & 0 & -\frac{A_{\bm k}^*}{c_3^*\sqrt{B_{\bm k}^2+|A_{\bm k}|^2}} & \frac{A_{\bm k}^*}{c_4^*\sqrt{B_{\bm k}^2+|A_{\bm k}|^2}} \\
0 & 0 & -\frac{B_{\bm k}+\sqrt{B_{\bm k}^2+|A_{\bm k}|^2}}{c_3^*\sqrt{B_{\bm k}^2+|A_{\bm k}|^2}} & \frac{B_{\bm k}-\sqrt{B_{\bm k}^2+|A_{\bm k}|^2}}{c_4^*\sqrt{B_{\bm k}^2+|A_{\bm k}|^2}}
\end{pmatrix}.
\end{align}
Here the coefficients $c_1$, $c_2$, $c_3$, and $c_4$ can be obtained by $X(\bm k)\cdot N\cdot X^{\dagger}(\bm k)=N$. Therefore we have
\begin{align}
\frac{1}{2|c_1|^2}=\frac{1}{2|c_3|^2}=\sqrt{B_{\bm k}^2+|A_{\bm k}|^2}\left(\sqrt{B_{\bm k}^2+|A_{\bm k}|^2}-B_{\bm k}\right), \\
\frac{1}{2|c_2|^2}=\frac{1}{2|c_4|^2}=\sqrt{B_{\bm k}^2+|A_{\bm k}|^2}\left(\sqrt{B_{\bm k}^2+|A_{\bm k}|^2}+B_{\bm k}\right), 
\end{align}
and finally the inverse of the paraunitary matrix Eq.~(\ref{SM_invpara}) becomes 
\begin{align}\label{inv_para}
	X^{-1}(\bm k)=\frac{1}{\sqrt{2}\Gamma^*_{\bm k}}
	\begin{pmatrix}
	\Gamma^*_{\bm k}K_{\bm k}^- & -\Gamma^*_{\bm k}K_{\bm k}^+ & 0 & 0 \\
	|\Gamma_{\bm k}|K_{\bm k}^+ & |\Gamma_{\bm k}|K_{\bm k}^- & 0 & 0 \\
	0 & 0 & \Gamma^*_{\bm k}K_{\bm k}^- & -\Gamma^*_{\bm k}K_{\bm k}^+ \\
	0 & 0 & |\Gamma_{\bm k}|K_{\bm k}^+ & |\Gamma_{\bm k}|K_{\bm k}^-
	\end{pmatrix},
	\end{align}
	where
	\begin{align}
	K_{\bm k}^{\pm}=\sqrt{1\pm\frac{B_{\bm k}}{\sqrt{B_{\bm k}^2+4|\Gamma_{\bm k}|^2}}}.
	\end{align}

As we mentioned in the main text, the magnon OAM is generally given by Eq.~(10)~\cite{SM_Fishman2022,SM_Fishman_review_2022,SM_Fishman2023}. By using the inverse of the paraunitary matrix $X^{-1}(\bm k)$, we can calculate the magnon OAM as
\begin{align}
\mathcal{L}^{\rm R}_{z1}(\bm k)&=\frac{\hbar}{2}\Bigg\{X^{-1}_{11}(\bm k)\hat{l}_{z\bm k}X^{-1}_{11}(\bm k)^*+X^{-1}_{21}(\bm k)\hat{l}_{z\bm k}X^{-1}_{21}(\bm k)^*\Bigg\} \\
&=\frac{\hbar}{2}\Bigg\{\frac{K_{\bm k}^-}{\sqrt{2}}\hat{l}_{z\bm k}\frac{K_{\bm k}^-}{\sqrt{2}}+\frac{K_{\bm k}^+}{\sqrt{2}}\frac{|\Gamma_{\bm k}|}{\Gamma^*_{\bm k}}\hat{l}_{z\bm k}\left(\frac{K_{\bm k}^+}{\sqrt{2}}\frac{|\Gamma_{\bm k}|}{\Gamma_{\bm k}}\right)\Bigg\} \\
&=\frac{\hbar}{4}\Bigg\{K_{\bm k}^+\hat{l}_{z\bm k}K_{\bm k}^++K_{\bm k}^-\hat{l}_{z\bm k}K_{\bm k}^-+(K_{\bm k}^+)^2\frac{\Gamma_{\bm k}}{|\Gamma_{\bm k}|}\hat{l}_{z\bm k}\frac{\Gamma^*_{\bm k}}{|\Gamma_{\bm k}|}\Bigg\} \\
&=\frac{\hbar}{4}\left(1+\frac{B_{\bm k}}{\sqrt{B_{\bm k}^2+4|\Gamma_{\bm k}|^2}}\right)\frac{\Gamma_{\bm k}}{|\Gamma_{\bm k}|}\hat{l}_{z\bm k}\frac{\Gamma^*_{\bm k}}{|\Gamma_{\bm k}|},
\end{align}
for the lower band, where $K_{\bm k}^+\hat{l}_{z\bm k}K_{\bm k}^++K_{\bm k}^-\hat{l}_{z\bm k}K_{\bm k}^-=0$. Similarly, the magnon OAM for the upper band can be calculated as
\begin{align}
\mathcal{L}^{\rm R}_{z2}(\bm k)=\frac{\hbar}{4}(K_{\bm k}^-)^2\frac{\Gamma_{\bm k}}{|\Gamma_{\bm k}|}\hat{l}_{z\bm k}\frac{\Gamma^*_{\bm k}}{|\Gamma_{\bm k}|}=\frac{\hbar}{4}\left(1-\frac{B_{\bm k}}{\sqrt{B_{\bm k}^2+4|\Gamma_{\bm k}|^2}}\right)\frac{\Gamma_{\bm k}}{|\Gamma_{\bm k}|}\hat{l}_{z\bm k}\frac{\Gamma^*_{\bm k}}{|\Gamma_{\bm k}|}.
\end{align}
On the other hand, we can also calculate the magnon OAM for the left-handed helix from the Hamiltonian Eq.~(\ref{SM_left}):
\begin{align}
\mathcal{L}^{\rm L}_{z1(2)}(\bm k)=\frac{\hbar}{4}(K_{\bm k}^{\pm})^2\frac{\Gamma^*_{\bm k}}{|\Gamma_{\bm k}|}\hat{l}_{z\bm k}\frac{\Gamma_{\bm k}}{|\Gamma_{\bm k}|}=\frac{\hbar}{4}\left(1\pm\frac{B_{\bm k}}{\sqrt{B_{\bm k}^2+4|\Gamma_{\bm k}|^2}}\right)\frac{\Gamma^*_{\bm k}}{|\Gamma_{\bm k}|}\hat{l}_{z\bm k}\frac{\Gamma_{\bm k}}{|\Gamma_{\bm k}|}.
\end{align}
Notice that
\begin{align}\label{Lz0gamma}
&\frac{\Gamma_{\bm k}}{|\Gamma_{\bm k}|}\hat{l}_{z\bm k}\frac{\Gamma^*_{\bm k}}{|\Gamma_{\bm k}|}=-\frac{\Gamma^*_{\bm k}}{|\Gamma_{\bm k}|}\hat{l}_{z\bm k}\frac{\Gamma_{\bm k}}{|\Gamma_{\bm k}|} \nonumber \\
=&\frac{-1}{|\gamma_{\bm k}|^2}\Bigg\{\sqrt{3}\bar{k}_x\sin{K_1}\sin{K_2}-\bar{k}_y\left[\left(\cos{K_1}+2\cos{K_2}\right)\left(\cos{K_1}-\cos{K_2}\right)+\sin^2{K_1}\right]\Bigg\},
\end{align}
with $K_1=3k_x/2$, $K_2=\sqrt{3}k_y/2$, and $\gamma_{\bm k}=\sqrt{1+4\cos{K_1}\cos{K_2}+4\cos^2{K_2}}$. Here, we introduce
\begin{align}
&\bar{k}_x=\sin{(3k_xa/2)}\cos{(\sqrt{3}k_ya/2)}, \nonumber \\ 
&\bar{k}_y=\frac{1}{\sqrt{3}a}\left\{\sin{(\sqrt{3}k_ya/2)}\cos{(3k_xa/2)}+\sin{(\sqrt{3}k_ya)}\right\},
\end{align}
corresponding to the two-dimensional (2D) honeycomb layders to guarantee the periodicity of magnon OAM on a discrete lattice~\cite{SM_Fishman2022,SM_Fishman_review_2022}.
Eq.~(\ref{Lz0gamma}) can have opposite sign if we choose the unit cells to be mirror symmetric as shown in Fig.~\ref{sm_struct}. Thus, we have $\mathcal{L}^{\rm L}_{z1(2)}(\bm k)=-\mathcal{L}^{\rm R}_{z1(2)}(\bm k)$ which reflects the chirality of the magnon OAM in chiral structures.

\section{Experimental proposal}
In this section, we provide an experimental proposal for magnon OAM detection.
First, we discuss the relation between the magnon OAM and Berry connection for our system. It has been clarified that the magnon OAM can have a topological nature~\cite{SM_Fishman2022,SM_Fishman_review_2022,SM_Fishman2023}, i.e, the magnon OAM $\mathcal{L}_{zn}(\bm k)$ can be rewritten by the cross product of the momentum $\bm k$ and the Berry connection $\bm{\mathcal{A}}_{n}(\bm k)$ as follows:
\begin{align}\label{Lz_Berry}
\mathcal{L}_{zn}(\bm k)=&-\frac{i\hbar}{2}\Big\{\bm k\times\braket{u_n(\bm k)|\nabla_{\bm k}u_n(\bm k)}\Big\}\cdot\hat{\bm z} \nonumber\\
=&-\frac{i\hbar}{2}\sum_{r=1}^{M}\Bigg\{\bm k\times\Bigg[X^{-1}_{rn}(\bm k)^*\frac{\partial X^{-1}_{rn}(\bm k)}{\partial\bm k}-X^{-1}_{r+M,n}(\bm k)^*\frac{\partial X^{-1}_{r+M,n}(\bm k)}{\partial\bm k}\Bigg]\Bigg\}\cdot\hat{\bm z} \nonumber\\
=&-\frac{\hbar}{2}\left[\bm k\times\bm{\mathcal{A}}_{n}(\bm k)\right]_z.
\end{align}
Here the Berry connection for the $n$-th magnon band is given by
\begin{align}\label{Berry_conn}
\bm{\mathcal{A}}_{n}(\bm k)\equiv i\sum_{r=1}^{M}\Bigg[X^{-1}_{rn}(\bm k)^*\frac{\partial X^{-1}_{rn}(\bm k)}{\partial\bm k}-X^{-1}_{r+M,n}(\bm k)^*\frac{\partial X^{-1}_{r+M,n}(\bm k)}{\partial\bm k}\Bigg].
\end{align}
By using Eq.~(\ref{inv_para}), we can calculate the Berry connection for the right-handed chiral crystal as
\begin{align}
\mathcal{A}^{R}_{\alpha1(2)}(\bm k)=&\frac{i}{2|\Gamma_{\bm k}|^2}\left(\Gamma_{\bm k}^*\frac{\partial\Gamma_{\bm k}}{\partial k_{\alpha}}-|\Gamma_{\bm k}|\frac{\partial|\Gamma_{\bm k}|}{\partial k_{\alpha}}\right)\left(1\pm\frac{B_{\bm k}}{\sqrt{B_{\bm k}+4|\Gamma_{\bm k}|^2}}\right) \nonumber\\
=&-\frac{1}{2}\frac{\partial\theta_{\bm k}}{\partial k_{\alpha}}\left(1\pm\frac{B_{\bm k}}{\sqrt{B_{\bm k}+4|\Gamma_{\bm k}|^2}}\right),
\end{align}
where $\alpha=x,y$, $\theta_{\bm k}=\text{arg}~\Gamma_{\bm k}$, and $\Gamma_{\bm k}$ and $B_{\bm k}$ are defined in the previous sections.
Here we notice that the first term
\begin{align}
\mathcal{A}^{R}_{\alpha0}(\bm k)=-\frac{1}{2}\frac{\partial\theta_{\bm k}}{\partial k_{\alpha}}
\end{align}
denotes the contribution from the 2D honeycomb layers, corresponding to $\mathcal{L}_0(\bm k)$. The second term
\begin{align}\label{2ndterm}
\Delta\mathcal{A}^{R}_{\alpha1(2)}(\bm k)=\mp\frac{1}{2}\frac{\partial\theta_{\bm k}}{\partial k_{\alpha}}\frac{B_{\bm k}}{\sqrt{B_{\bm k}+4|\Gamma_{\bm k}|^2}}
\end{align}
contributes to $\Delta\mathcal{L}^{R}_{z1(2)}(\bm k)$, which comes from the chiral exchange interaction as discussed in the main text.
On the other hand, we can also calculate the Berry connection for the left-handed chiral crystal as
\begin{align}
\mathcal{A}^{L}_{\alpha1(2)}(\bm k)=&\frac{i}{2|\Gamma_{\bm k}|^2}\left(\Gamma_{\bm k}\frac{\partial\Gamma^*_{\bm k}}{\partial k_{\alpha}}-|\Gamma_{\bm k}|\frac{\partial|\Gamma_{\bm k}|}{\partial k_{\alpha}}\right)\left(1\pm\frac{B_{\bm k}}{\sqrt{B_{\bm k}+4|\Gamma_{\bm k}|^2}}\right) \nonumber\\
=&\frac{1}{2}\frac{\partial\theta_{\bm k}}{\partial k_{\alpha}}\left(1\pm\frac{B_{\bm k}}{\sqrt{B_{\bm k}+4|\Gamma_{\bm k}|^2}}\right)=-\mathcal{A}^{R}_{\alpha1(2)}(\bm k).
\end{align}
The signs of the Berry connection between the left- and right-handed chiral crystals are opposite.

% Fig.S2	
	\begin{figure}[htb]
    %\begin{figure*}[htb]
    \begin{center}
    \includegraphics[clip,width=15cm]{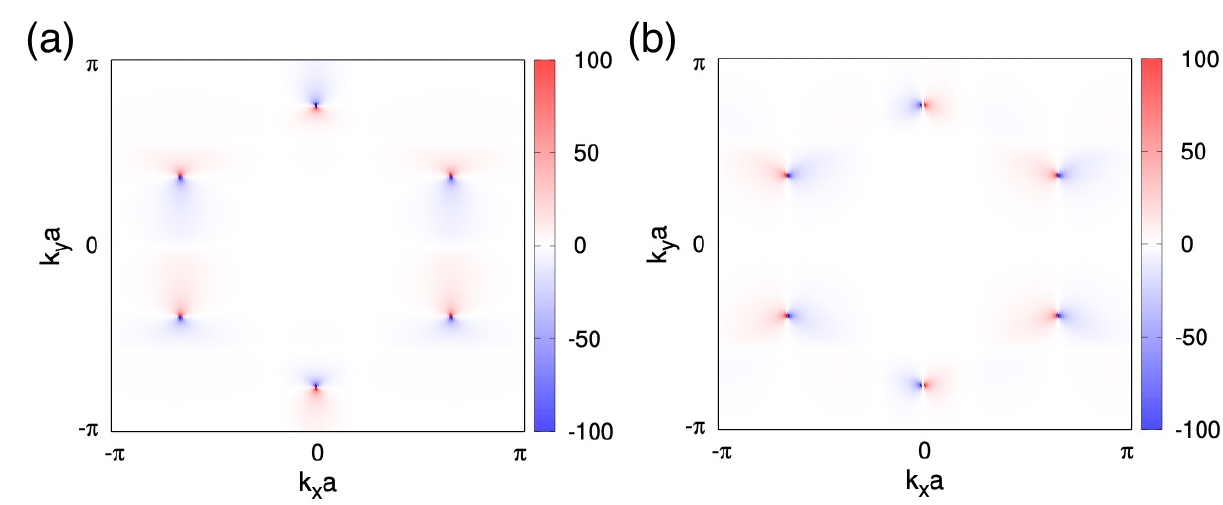}
    \end{center}
    \caption{Distributions of $\Delta\mathcal{A}_{\alpha2}(\bm k)$ for the second magnon band in the $k_x$-$k_y$ plane with $k_z=\pi/3$.
    Colors represent $\Delta\mathcal{A}_{x2}(\bm k)$ in (a) and $\Delta\mathcal{A}_{y2}(\bm k)$ in (b), respectively. The parameters are set to be $J_2=0.5J_1$.
    }
     \label{sm_Berry}
     \end{figure}
     %\end{figure*}	

\begin{figure}[htb]
    %\begin{figure*}[htb]
    \begin{center}
    \includegraphics[clip,width=15cm]{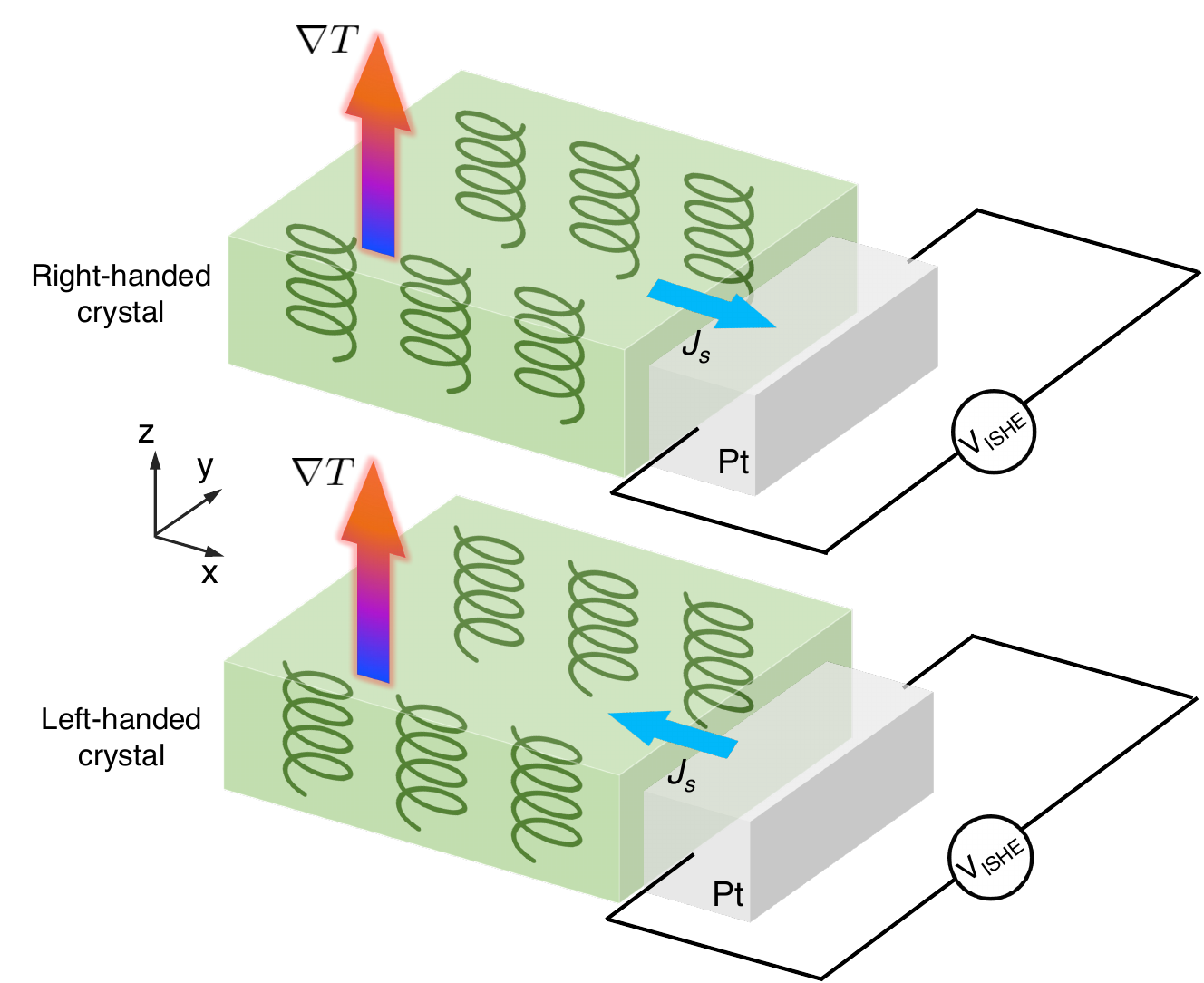}
    \end{center}
    \caption{Schematic illustration of the magnon OAM detection by utilizing the inverse spin Hall effect (ISHE) in Pt.
    }
     \label{sm_exp}
     \end{figure}

Next, we discuss experimental proposal for the magnon OAM detection. We focus on the right-handed chiral crystal as an example in the following calculations. Since $\Delta\mathcal{A}_{\alpha1(2)}(\bm k)$ plays an essential role in the magnon orbital Edelstein effect (MOEE), here we show the distributions of $\Delta\mathcal{A}_{x2}(\bm k)$ and $\Delta\mathcal{A}_{y2}(\bm k)$ in the $k_x$-$k_y$ plane with $k_z=\pi/3$ in Fig.~\ref{sm_Berry}(a) and (b), respectively. We find that the Berry connections show sharp peaks near the valley points $\bm K$ and $\bm K'$ and plummets suddenly away from these points.
In this case, the MOEE coefficient $\alpha_{zz}$ given by Eq.~(18) in the main text can be approximately expressed as
\begin{align}
\alpha_{zz}=&-\frac{2\tau}{\hbar}\sum_{n}\int\frac{d\bm k}{(2\pi)^{3}}\Delta\mathcal{L}_{zn}(\bm k)\frac{\partial E_{n\bm k}}{\partial k_z}\frac{\partial f_0(E_{n\bm k})}{\partial T} \nonumber\\
=&~\tau\sum_{n}\int\frac{d\bm k}{(2\pi)^3}\left[\bm k\times\Delta\bm{\mathcal{A}}_{n}(\bm k)\right]_z\frac{\partial E_{n\bm k}}{\partial k_z}\frac{\partial f_0(E_{n\bm k})}{\partial T} \nonumber\\
\approx&~\tau\sum_{n}\left[(\bm K+\bm K')\times\int\frac{d\bm k}{(2\pi)^{3}}\Delta\bm{\mathcal{A}}_{n}(\bm k)\frac{\partial E_{n\bm k}}{\partial k_z}\frac{\partial f_0(E_{n\bm k})}{\partial T}\right]_z \nonumber\\
=&~2\tau\left[\bm K\times\Delta\bm P_{\text{mag}}\right]_z,
\end{align}
where
\begin{align}
\Delta\bm P_{\text{mag}}=\sum_{n}\int\frac{d\bm k}{(2\pi)^{3}}\Delta\bm{\mathcal{A}}_{n}(\bm k)\frac{\partial E_{n\bm k}}{\partial k_z}\frac{\partial f_0(E_{n\bm k})}{\partial T}
\end{align}
represents the change of the magnon ``polarization", corresponding to the change of magnon positions in analogy to electric polarization~\cite{SM_King-Smith1993}. Therefore, in the presence of the temperature gradient $\partial_zT$, the generated total magnon OAM in $z$ direction is accompanied by $\Delta\bm P_{\text{mag}}$ within the $xy$ plane. Then, the magnon (polarization) current flows perpendicularly to the screw axis, which can be detected via the inverse spin Hall effect as shown in Fig.~\ref{sm_exp}. In this manner, the MOEE can be detected through the inverse spin Hall effect.

% Fig.S2	
     %\end{figure*}	

\section{Gauge invariance}
The magnon OAM with wave vector $\bm k$ for band $n$ is not gauge invariant while the gauge-invariance can be recovered by integrating the magnon OAM over the orientation $\phi$ of wave vector $\bm k$~\cite{SM_Fishman2023}.
In this section, we discuss that the magnon OAM is gauge invariant under a gauge transformation dependent only on $k=|\bm k|$. We preform the following gauge transformation
\begin{align}
X^{-1}_{rn}(\bm k)\rightarrow X^{-1}_{rn}(\bm k)e^{-i\lambda_n(k)},
\end{align}
where $\lambda_n(k)$ depends only on $k$, and then the gradient of $X^{-1}_{rn}(\bm k)$ in $\bm k$ space becomes
\begin{align}
\frac{\partial X^{-1}_{rn}(\bm k)}{\partial\bm k}\rightarrow&\frac{\partial X^{-1}_{rn}(\bm k)}{\partial\bm k}e^{-i\lambda_n(k)}-i\frac{\partial\lambda_n(k)}{\partial\bm k}X^{-1}_{rn}(\bm k)=\frac{\partial X^{-1}_{rn}(\bm k)}{\partial\bm k}e^{-i\lambda_n(k)}-i\frac{\bm k}{k}\frac{\partial\lambda(k)}{\partial k}X^{-1}_{rn}(\bm k).
\end{align}
Thus, the Berry connection given by Eq.~(\ref{Berry_conn}) under this gauge transformation becomes
\begin{align}
\bm{\mathcal{A}}_n(\bm k)\rightarrow\bm{\mathcal{A}}_n(\bm k)+\frac{\bm k}{k}\frac{\partial\lambda_n(k)}{\partial k}e^{i\lambda_n(k)}\sum_{r=1}^{M}\left(|X^{-1}_{rn}(\bm k)|^2-|X^{-1}_{r+M,n}(\bm k)|^2\right),
\end{align}
where 
\begin{align}
\sum_{r=1}^{M}\left(|X^{-1}_{rn}(\bm k)|^2-|X^{-1}_{r+M,n}(\bm k)|^2\right)=1,
\end{align}
because of the normalization condition $X(\bm k)\cdot N\cdot X^{\dagger}(\bm k)=N$.
Therefore, we can show that the magnon OAM in Eq.~(\ref{Lz_Berry}) is gauge invariant under this gauge transformation:
\begin{align}
\mathcal{L}_{zn}(\bm k)\rightarrow\mathcal{L}_{zn}(\bm k).
\end{align}

\section{Effectiveness of the lattice model}
\begin{figure}[htb]
    %\begin{figure*}[htb]
    \begin{center}
    \includegraphics[clip,width=15cm]{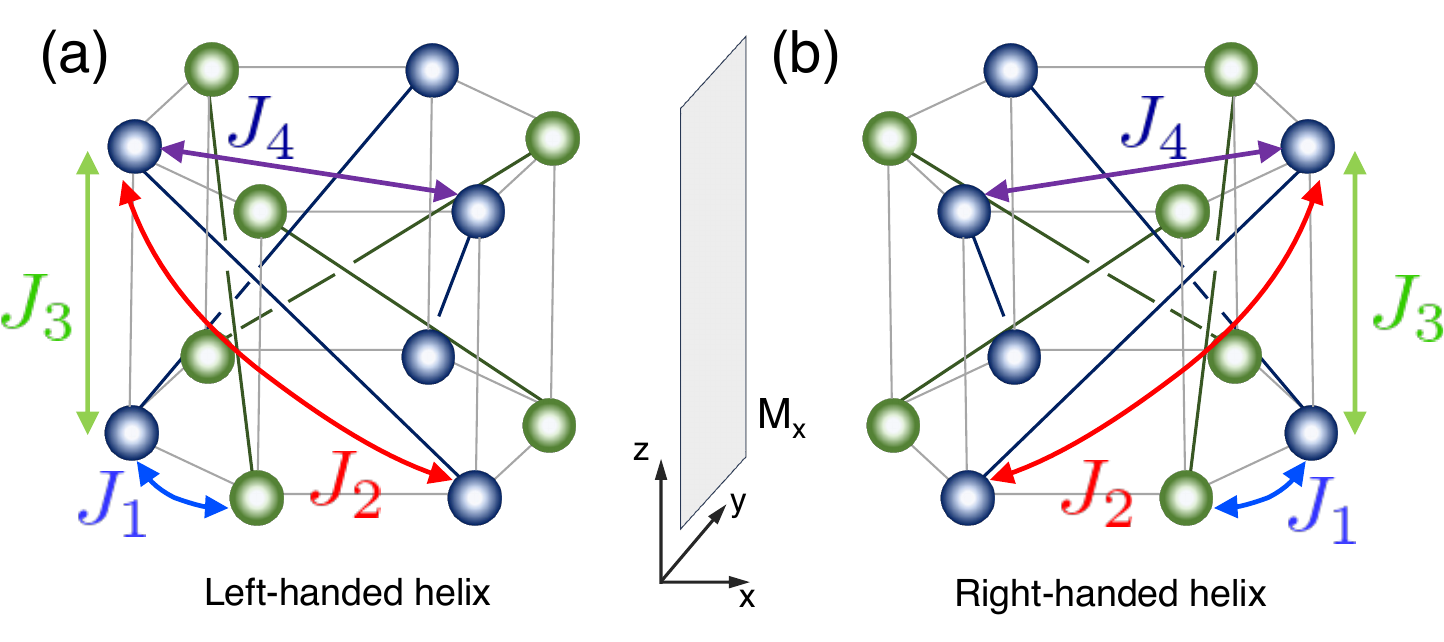}
    \end{center}
    \caption{Chiral crystals composed of a stacked honeycomb lattice with (a) left-handed helix and (b) right-handed helix. Here $J_1>0$ is the nearest-neighbor exchange interaction within the $xy$ plane, $J_2>0$ is the chiral exchange interaction between two layers, $J_3$ is the interlayer exchange interaction along the bonds in a non-chiral way, and $J_4$ is the next-nearest-neighbor exchange interaction within the $xy$ plane.
    }
     \label{sm_eff}
     \end{figure}

Our model is intended to be a minimal model to demonstrate the chiral magnon with magnon OAM in ferromagnetic chiral crystals. In this section, we discuss the effectiveness of the lattice model, and analyze the result including a few more interaction terms. We consider the third and fourth interaction terms in the Heisenberg spin Hamiltonian with the exchange interactions $J_3$ and $J_4$ as shown in Fig.~\ref{sm_eff}.

% Fig.S2	
     %\end{figure*}	
     
Now, we write the total spin Hamiltonian as
\begin{align}
H=-J_1\sum_{\braket{ij}}\bm S_i\cdot\bm S_j-J_2\sum_{[ij]}\bm S_i\cdot\bm S_j-J_3\sum_{\{ij\}}\bm S_i\cdot\bm S_j-J_4\sum_{\braket{\braket{ij}}}\bm S_i\cdot\bm S_j,
\end{align}
where $\{ij\}$ represents summation over the sites along the bonds in the $z$ direction, and $\braket{\braket{ij}}$ denotes the summation over the next-nearest-neighbor sites within the $xy$ plane.
By means of the Holstein-Primakoff transformation~\cite{SM_HP1940}, the spin Hamiltonian is rewritten as
\begin{align}
H=&-J_1S\sum_{\braket{ij}}(a_i^{\dagger}a_j+\text{h.c})-J_2S\sum_{[ij]}(a_i^{\dagger}a_j+\text{h.c})-J_3S\sum_{\{ij\}}(a_i^{\dagger}a_j+\text{h.c}) \nonumber\\
&-J_4S\sum_{\braket{\braket{ij}}}(a_i^{\dagger}a_j+\text{h.c})+(3J_1+6J_2+2J_3+6J_4)S\sum_ia_i^{\dagger}a_i.
\end{align}
Then the total Bloch Hamiltonian including the third and fourth interaction terms is given by
\begin{align}
\mathcal{H}_{\text{tot}}(\bm k)=\mathcal{H}^{\alpha}(\bm k)+\mathcal{H}'(\bm k),
\end{align} 
where $\alpha=R,L$, $\mathcal{H}^{R}(\bm k)$ and $\mathcal{H}^{L}(\bm k)$ are given by Eqs.~(\ref{SM_right}) and (\ref{SM_left}), and
\begin{align}
\mathcal{H}'(\bm k)=\frac{1}{2}
\begin{pmatrix}
h'(\bm k) & 0 & 0 & 0 \\
0 & h'(\bm k) & 0 & 0 \\
0 & 0 & h'(\bm k) & 0 \\
0 & 0 & 0 & h'(\bm k) \\
\end{pmatrix}
\end{align}
with
\begin{align}
h'(\bm k)=2J_3S\Bigg\{1-\cos{(k_zc)}\Bigg\}+6J_4S\Bigg\{\cos{\left(\sqrt{3}k_ya\right)}+2\cos{\left(\frac{3k_xa}{2}\right)}\cos{\left(\frac{\sqrt{3}k_ya}{2}\right)}\Bigg\}.
\end{align}
Here we notice that the contributions from the third and fourth terms with the exchange interactions $J_3$ and $J_4$ are proportional to the identity matrix. Therefore they do not change the eigenstates of magnons.
On the other hand, the contributions from the first and second interaction terms play an important roles, resulting in the opposite sign of the magnon OAM for left- and right-handed chiral crystals.
Hence, the magnon OAM is unchanged even though a few more interaction terms in the lattice model are taken into account, and the conclusion also remains the same.